\documentclass{emulateapj}
\usepackage{comment}
\usepackage{epstopdf}
\begin{document}

\title{Complexity on dwarf galaxy scales: A bimodal distribution function in Sculptor}

\author{Maarten A. Breddels
  and
  Amina Helmi
}
\affil{Kapteyn Astronomical Institute, University of Groningen}
\affil{P.O. Box 800, 9700 AV Groningen, The Netherlands}
\email{breddels@astro.rug.nl}

\newcommand{\Ei}{\ensuremath{{\rm{E}_i}}}
\newcommand{\Lj}{\ensuremath{{\rm{L}_j}}}
\newcommand{\vlos}{\ensuremath{{\rm{v}_\text{los}}}}
\newcommand{\vlosi}{\ensuremath{{\rm{v}_\text{los,i}}}}

\newcommand{\params}{\ensuremath{{\boldsymbol{\theta}}}}
\newcommand{\weights}{\ensuremath{{\mathbf{w}}}}
\newcommand{\Rell}{\ensuremath{{R_\epsilon}}}
\newcommand{\Relli}{\ensuremath{{R_{\epsilon,i}}}}

\newcommand{\diff}{{\rm{d}}}
\newcommand{\myvec}[1]{{\boldsymbol{#1}}}
\newcommand{\dm}{\ensuremath{{\rm{DM}}}}
\newcommand{\true}{\ensuremath{{\rm{true}}}}
\newcommand{\maxt}{\ensuremath{{\rm{max}}}}
\newcommand{\mint}{\ensuremath{{\rm{min}}}}
\newcommand{\model}{\ensuremath{{\rm{model}}}}
\newcommand{\light}{\ensuremath{{\rm{light}}}}
\newcommand{\data}{\ensuremath{{\rm{data}}}}
\newcommand{\kin}{\ensuremath{{\rm{kin}}}}
\newcommand{\nudm}{\ensuremath{\nu_\dm}}
\newcommand{\Msol}{\ensuremath{\text{M}_{\odot}}}
\newcommand{\Lsol}{{\rm L}_{\odot}}
\newcommand{\Mkpc}{\ensuremath{\text{M}_{1\text{kpc}}}\xspace}
\newcommand{\los}{\ensuremath{\text{los}}\xspace}
\newcommand{\commenttext}[1]{({\it comment: {#1}})}
\newcommand{\TODO}[1]{({\bf TODO: {#1}})}
\newcommand{\NOTE}[1]{({\bf NOTE: {#1}})}
\newcommand{\reftodo}[1]{({\it TODO ref: {#1}})}
\newcommand{\maybe}[1]{({\it maybe: {#1}})}
\newcommand{\footnotecomment}[1]{\footnote{\it comment: {#1}}}
\newcommand{\kms}{\ensuremath{{\rm km\,s}^{-1}}\xspace}
\newcommand{\Lmax}{\ensuremath{L_\text{max}}}

\begin{abstract}
In previous work we have presented Schwarzschild models of the
Sculptor dSph, demonstrating that this system could be embedded in
dark matter halos that are either cusped or cored. Here we show that
the non-parametric distribution function recovered through
Schwarschild's method is bimodal in energy
and angular momentum space for all best fitting mass models explored. We demonstrate that this bimodality is
directly related to the two components known to be present in Sculptor through
stellar populations analysis, although our method is purely dynamical
in nature and does not use this prior information. It therefore
constitutes independent confirmation of the existence of two
physically distinct dynamical components in Sculptor and suggests a rather complex assembly history for this dwarf galaxy.

\end{abstract}


\section{Introduction}

Sculptor was the first dwarf spheroidal (dSph) galaxy discovered
\citep{Shapley1938}, and it may be considered an archetype of the class of
dSph as it is a rather featureless system \citep[compared to
e.g. Fornax,][]{deBoer2011AA}.  Nonetheless, Scl has proven to be more complex than
originally thought. Photometric surveys starting from
\cite{DaCosta1984}, \citep[and later by][]{Light1988PhDT,
  HurleyKeller1999ApJ, Majewski1999ApJ}, have revealed that its
horizontal branch morphology changes with radius. More recently,
thanks to large spectroscopic surveys, it has been possible to relate
the differences in the spatial distribution of the blue and red
horizontal branch (BHB/RHB) to differences in chemical composition and
kinematics of its stars \citep{Tolstoy2004ApJ}, as well as to age gradients \citep{deBoer2011AA}.

The picture that has emerged from this body of work is that Sculptor has two populations or components. The
first population is centrally concentrated, metal rich, younger, is
represented in RHB stars and has colder kinematics, with a decreasing
line-of-sight velocity dispersion with distance from the centre. The
second population is less concentrated, metal poor, older, is prominent in BHB stars, has hotter kinematics and a more
constant velocity dispersion profile. It was not clear until now whether this could simply be due to a population gradient.

dSph have been the target of many kinematic surveys in the past decade
because of their very high dynamical mass-to-light ratios.  The aim of these
studies has been to provide constraints on the nature and distribution
of dark matter, however no firm conclusions have yet been drawn. This is because of limitations
in the data (only access to line-of-sight velocities) and also in the models. For example, the
most-widely used modeling technique is based on the Jeans Equations and requires making assumptions
about the orbital structure of the system \citep[see
e.g.][]{Walker2009}, although recent work by
\citet{Richardson2013,Richardson2014} using higher moments and the
virial equations, seems to be able to circumvent this degeneracy. A
more powerful approach is to use orbit-based dynamical modeling, also
known as Schwarzschild's method. \cite{Breddels2012arXiv} have shown
that the data on Scl does not constrain the inner slope of its dark matter density
profile very strongly, and that neither cored nor cuspy \cite[of the NFW-type,][]{NFW1996ApJ...462..563N} profiles are favoured.

On the other hand, it has been argued that the multiple components
present in Sculptor should be used to model this system dynamically and that this
ought to lead to much tighter constraints. This is because these
components are hosted by the same underlying potential and their
presence effectively would reduce the available parameter space of
plausible dynamical models \citep{Battaglia2008ApJ}. Several studies
have attempted this using e.g. the virial equations \citep{Agnello} or
constraints based on Jeans modeling \citep{Walker2011ApJ,
  Amorisco2011}, and concluded that NFW-like profiles are strongly
disfavoured. All these works \citep[with the exception of ][]{Walker2011ApJ} have assumed that the two populations
are split in the same way in photometry and kinematics/chemistry,
although this is a priori not guaranteed, as the kinematics (and split
in metallicity) are obtained from the RGB while the photometry is fit
for the BHB and RHB stars independently. Perhaps more importantly,
all these models based on a single estimate of the mass at a given radius
(e.g. either through Jeans or through the virial equations) have
assumed the components to follow the same functional form of the light
distribution (with different characteristic parameters). As we shall
show below, this is not necessarily a valid asssumption.

In this {\it Letter} we analyse the phase-space structure of the best
fitting Schwarzschild dynamical models of Sculptor from
\citet{Breddels2013arXivb} and \citet{BreddelsThesis}. Our spherical orbit
based dynamical models are non-parametric in the distribution function
(e.g. no assumptions are made on the anisotropy) and
provide good fits to the global kinematics and light distribution of Scl.
The dark matter distribution follows specific parametric profiles
\citep[c.f.][]{Jardel2012Dra}, which all produce very similar mass distributions in a finite
region around the half-light radius of Scl. As we shall show below,
the orbit weights (which correspond to the distribution function of the galaxy) show a bimodal
distribution for all the best fit models of Sculptor, even though this is not a priori assumed.

\section{Distribution functions obtained via Schwarzschild models}

Our orbit based dynamical models are described extensively in
\citet{Breddels2012arXiv}. This method essentially consists in finding
a linear combination of orbits (integrated in a given gravitational
potential) that allows fitting the kinematic data and light
distribution of the system. By varying the characteristic parameters
of the gravitational potential (e.g. mass and scale radius), best fits
are found, while the weights of the orbits effectively provide
the distribution function non-parametrically.

\begin{figure}
\includegraphics[scale=0.45]{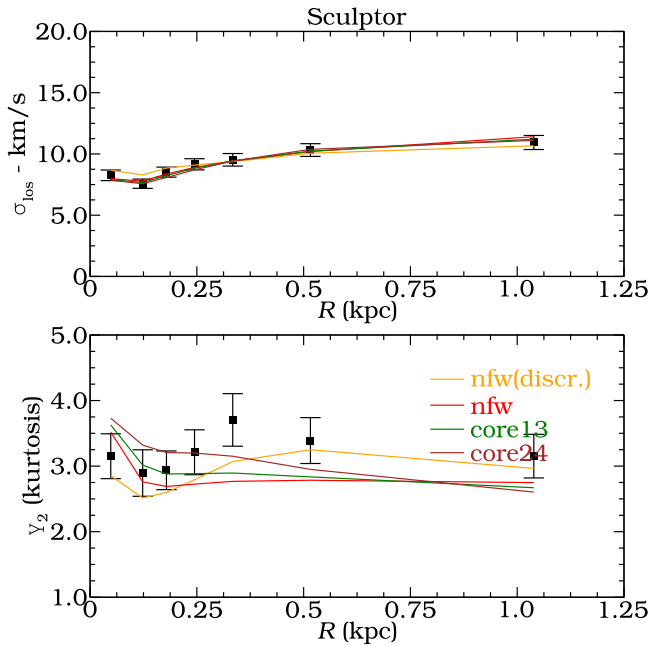}
\caption{Velocity dispersion (top panel) and kurtosis (lower panel) for the best fitting models for different dark matter profiles,
namely Navarro, Frenk \& White (NFW; based on moments or discrete modeling of the data) and
two cored profiles, with $\rho(r) \propto 1/(r + r_c)^3$ (core13) and $\rho(r) \propto 1/(r^2 + r_c^2)^2$ (core24).
\label{fig:vdispfit} }
\end{figure}

Fig.~\ref{fig:vdispfit} shows how well the best fit models for
various dark matter profiles explored in \citet{Breddels2013arXivb}
and from the discrete modeling approach by Breddels (2013, PhD thesis)
reproduce the velocity dispersion and kurtosis profiles.  As can be
seen from this figure the resulting fits are nearly indistinguishable
from one another.
\newcommand{\scaledf}{0.365}
\begin{figure}
\begin{tabular}{c c}
\textbf{discrete, nfw} & \textbf{moments, nfw} \\
\includegraphics[scale=\scaledf]{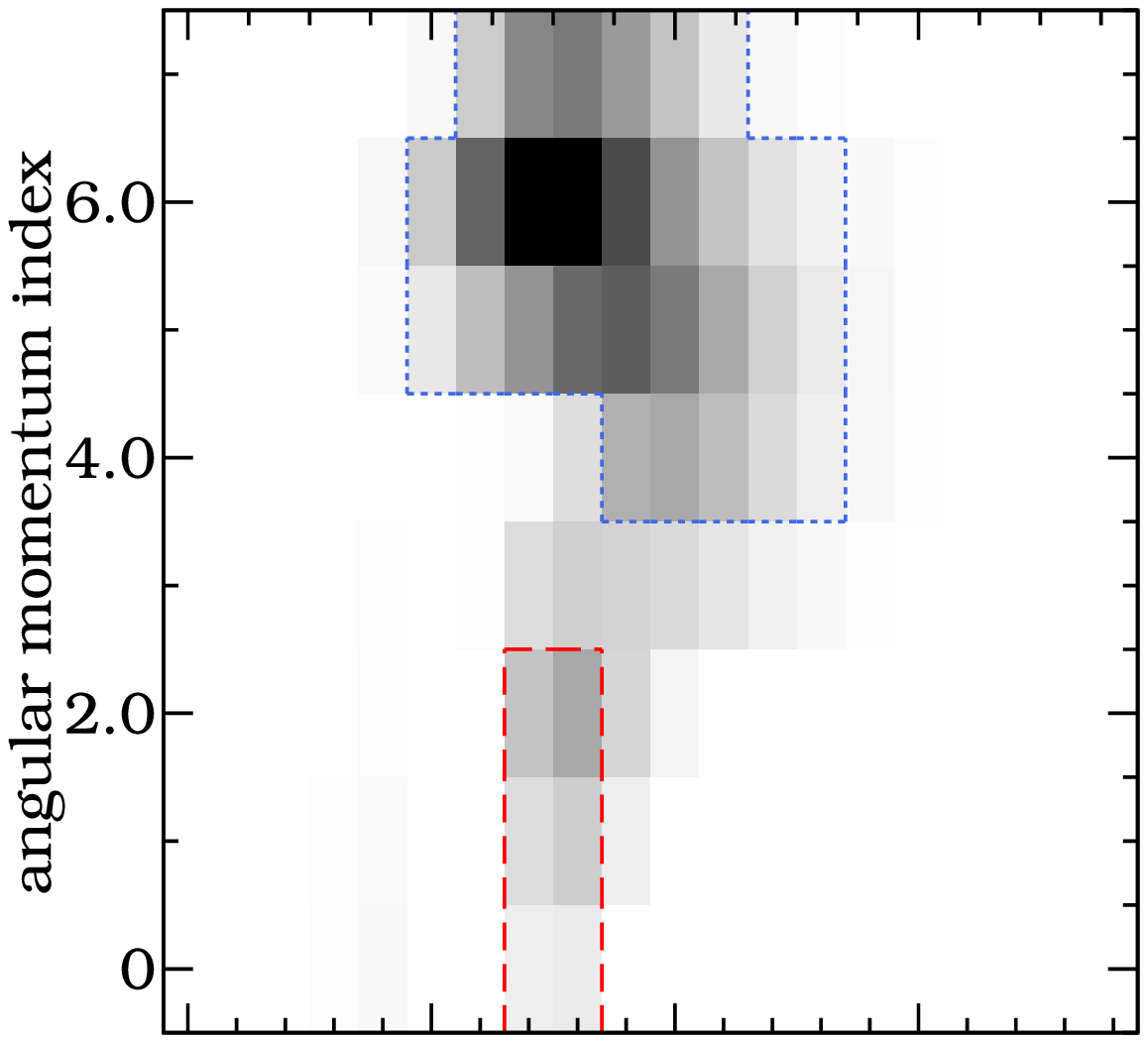} & \includegraphics[scale=\scaledf]{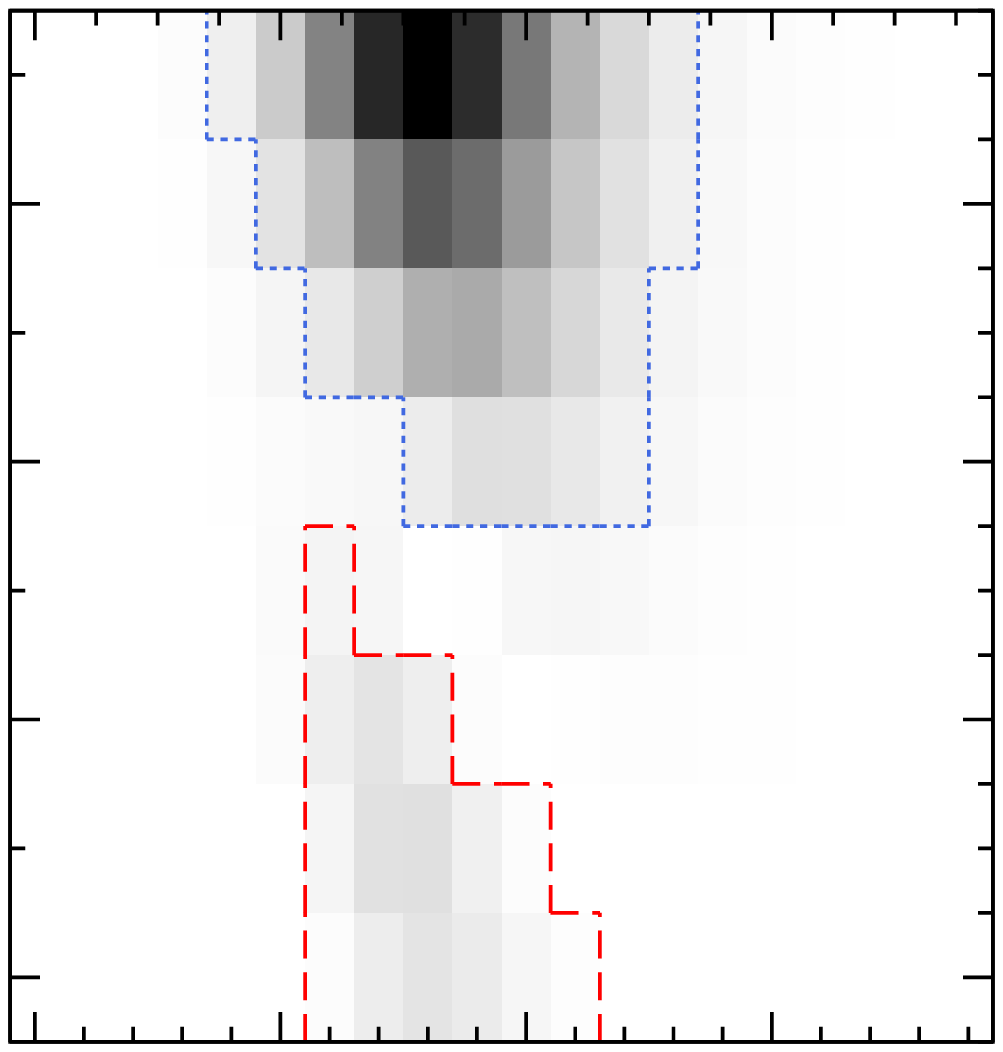}\\
\textbf{moments, core13} & \textbf{moments, core24} \\
\includegraphics[scale=\scaledf]{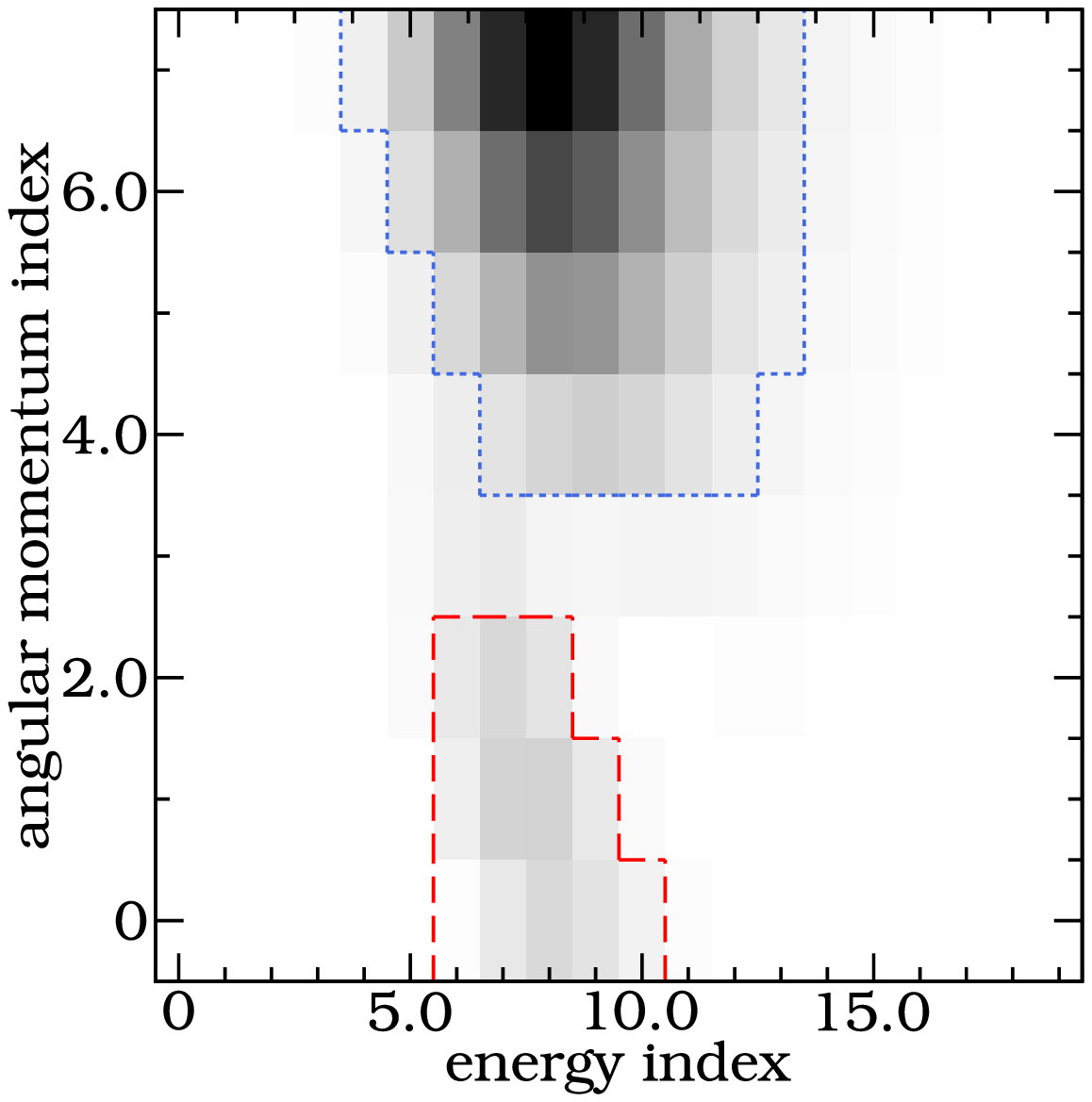} & \includegraphics[scale=\scaledf]{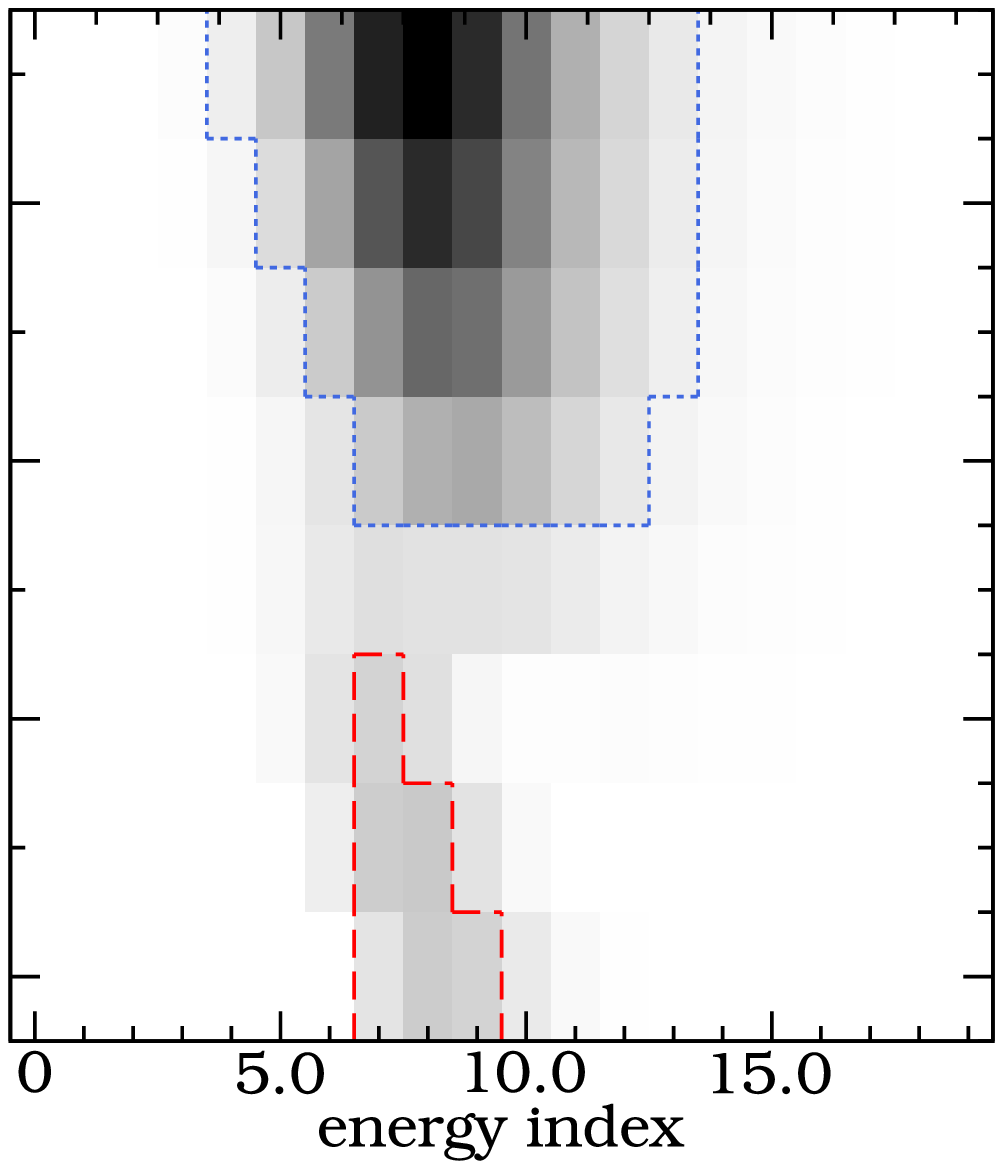}
\end{tabular}
\caption{Orbit-weights for the best fit models for different dark matter halo profiles: discrete
NFW \citep[][top left]{BreddelsThesis}; moment-based NFW (top right) and cored (bottom) models from \citet{Breddels2013arXivb}.
The red dashed and blue solid contours indicate the two components found by the watershed method.\label{fig:scl-df}}
\end{figure}

Fig.~\ref{fig:scl-df} shows the orbit-weights that define the
distribution functions of the best fit models explored.  For example,
the best fit NFW model found in Breddels (2013, PhD thesis),
obtained using discrete fitting is given in the upper left panel. The associated
distribution function appears to be bimodal as it has two distinct
components: one at low angular momentum (near energy index 8, angular
momentum index 2), and a second one at higher angular momentum (near
energy index 7, angular momentum index 6). This bimodality is also
present in the distribution functions of the remaining best fitting
models from \citet{Breddels2013arXivb}, whether cored or
cusped. Although the exact location of the lower angular momentum component varies slightly,
 it is reassuring that the bimodality is found using both the discrete and the traditional
Schwarzschild methods and also for different dark matter profiles. This demonstrates the the power of using non-parametric Schwarzschild methods.

To obtain further insights into the nature of these two components we
use a watershed algorithm. We first locate the local maxima in the
image, which serve to identify the two different components. Starting
from the pixel with the maxima, we follow the landscape downwards,
i.e. we find those neighbouring pixels that have a lower value. These
are associated to either of the components, until no more such pixels are
found. At this point we exclude pixels that could be simultaneously
associated to both components.  In Fig. \ref{fig:scl-df} we indicate
the pixels associated to each of the two components by showing the
contours that correspond to 5\% of the maximum
orbital-weight for each component. We then construct two distribution
functions by selecting the pixels identified by the
algorithm as just described. We now refer to the higher
and lower angular momentum components as the blue and red
components respectively.

\newcommand{\scalevdisp}{0.37}
\begin{figure}
\begin{tabular}{c c}
\textbf{discrete, nfw} & \textbf{moments, nfw} \\
\includegraphics[scale=\scalevdisp]{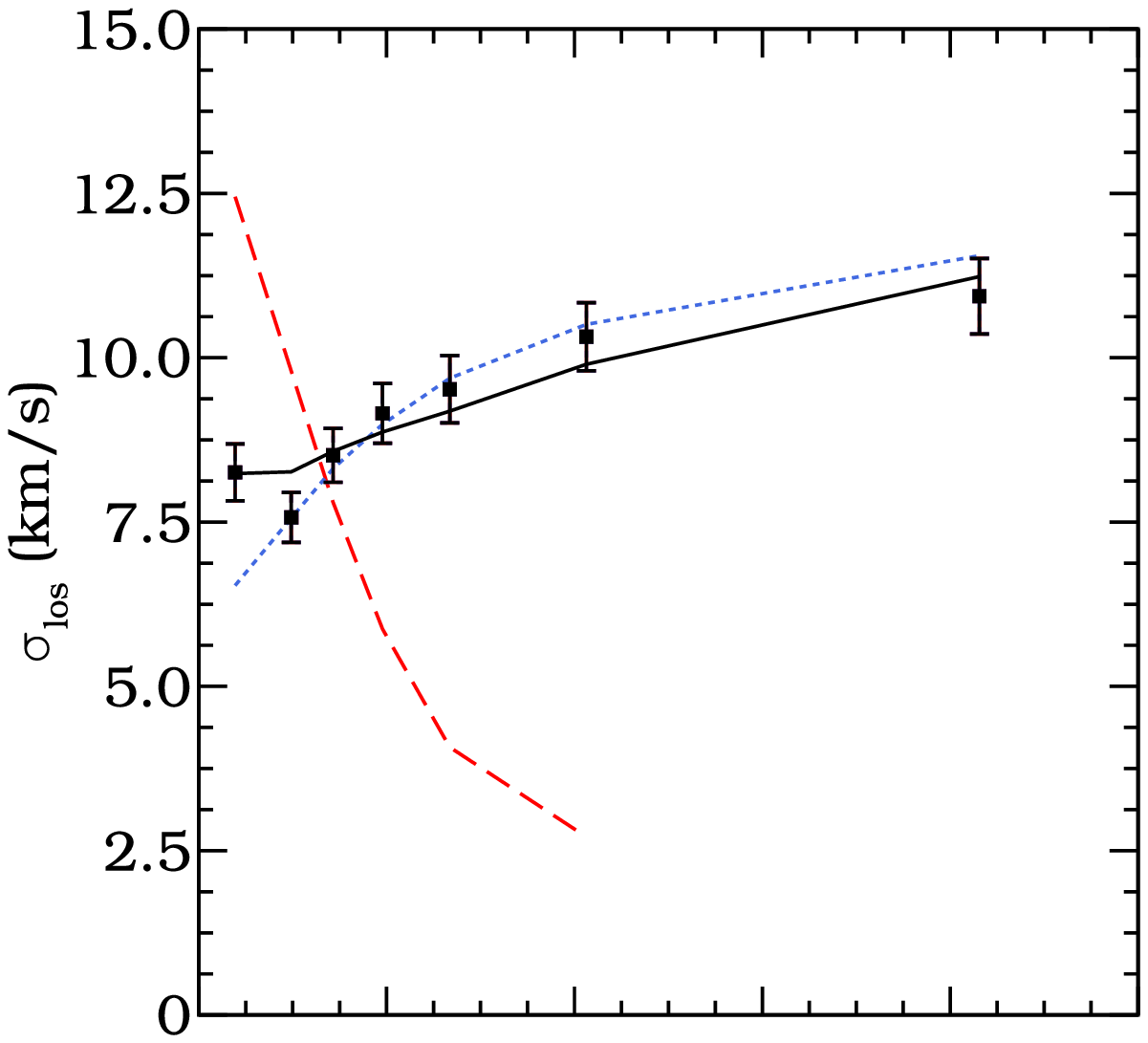} & \includegraphics[scale=\scalevdisp]{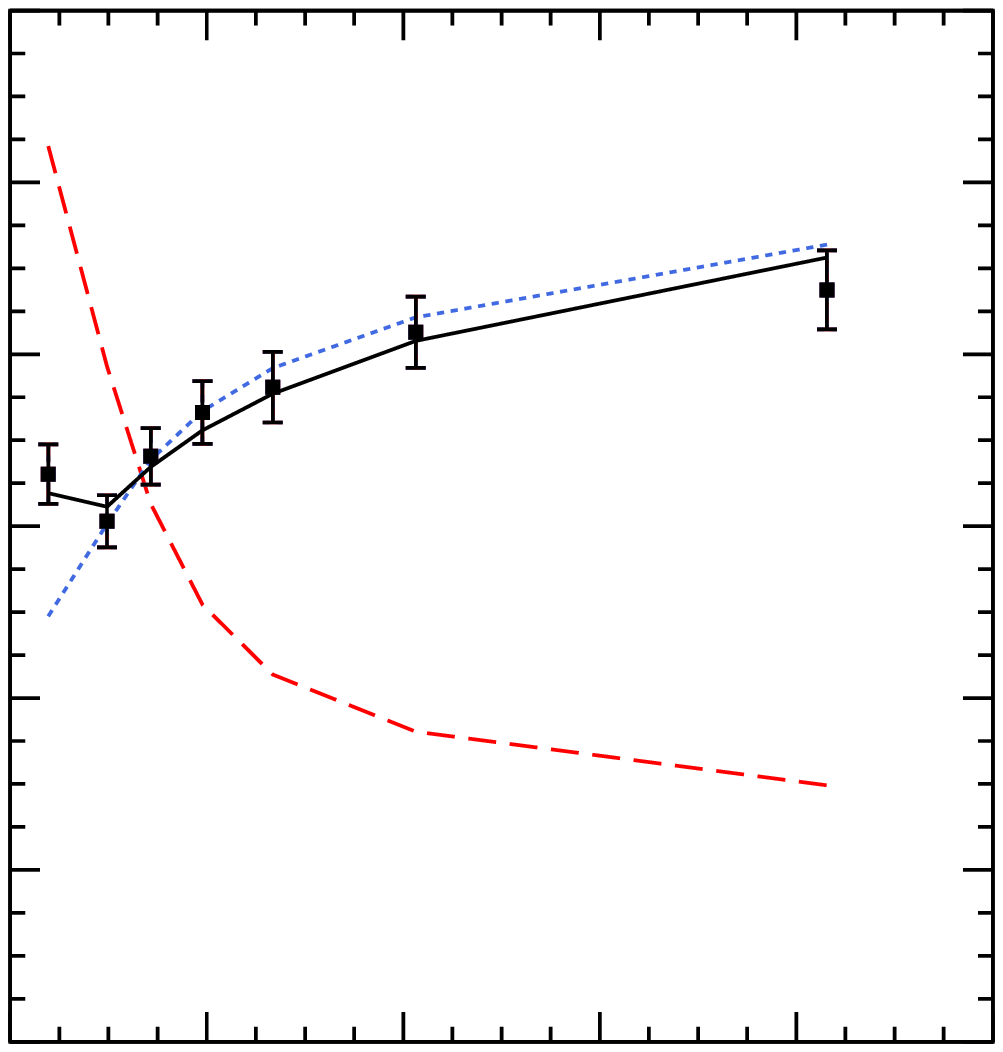} \\
  \textbf{moments, core13} & \textbf{moments, core24} \\
\includegraphics[scale=\scalevdisp]{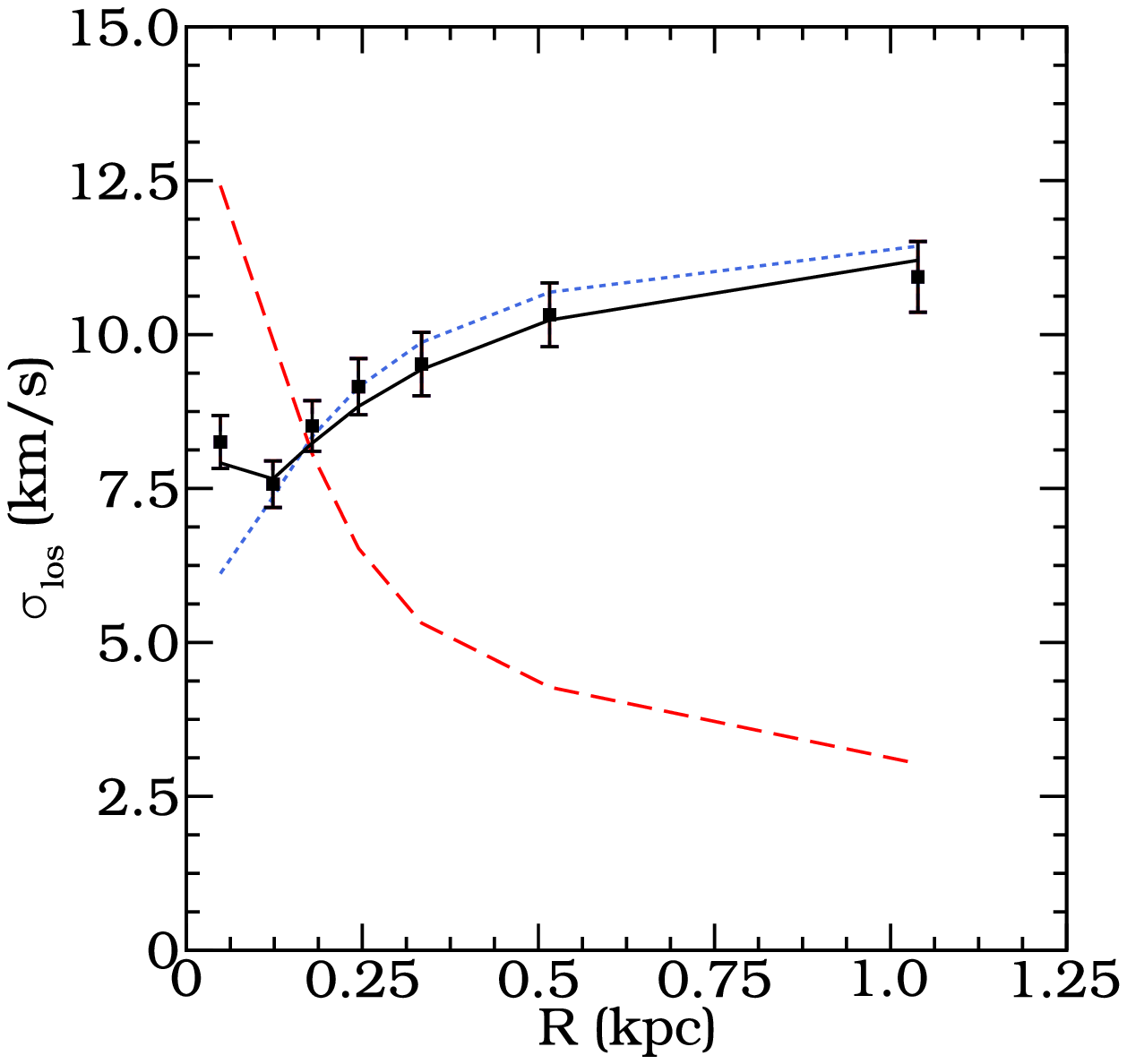} & \includegraphics[scale=\scalevdisp]{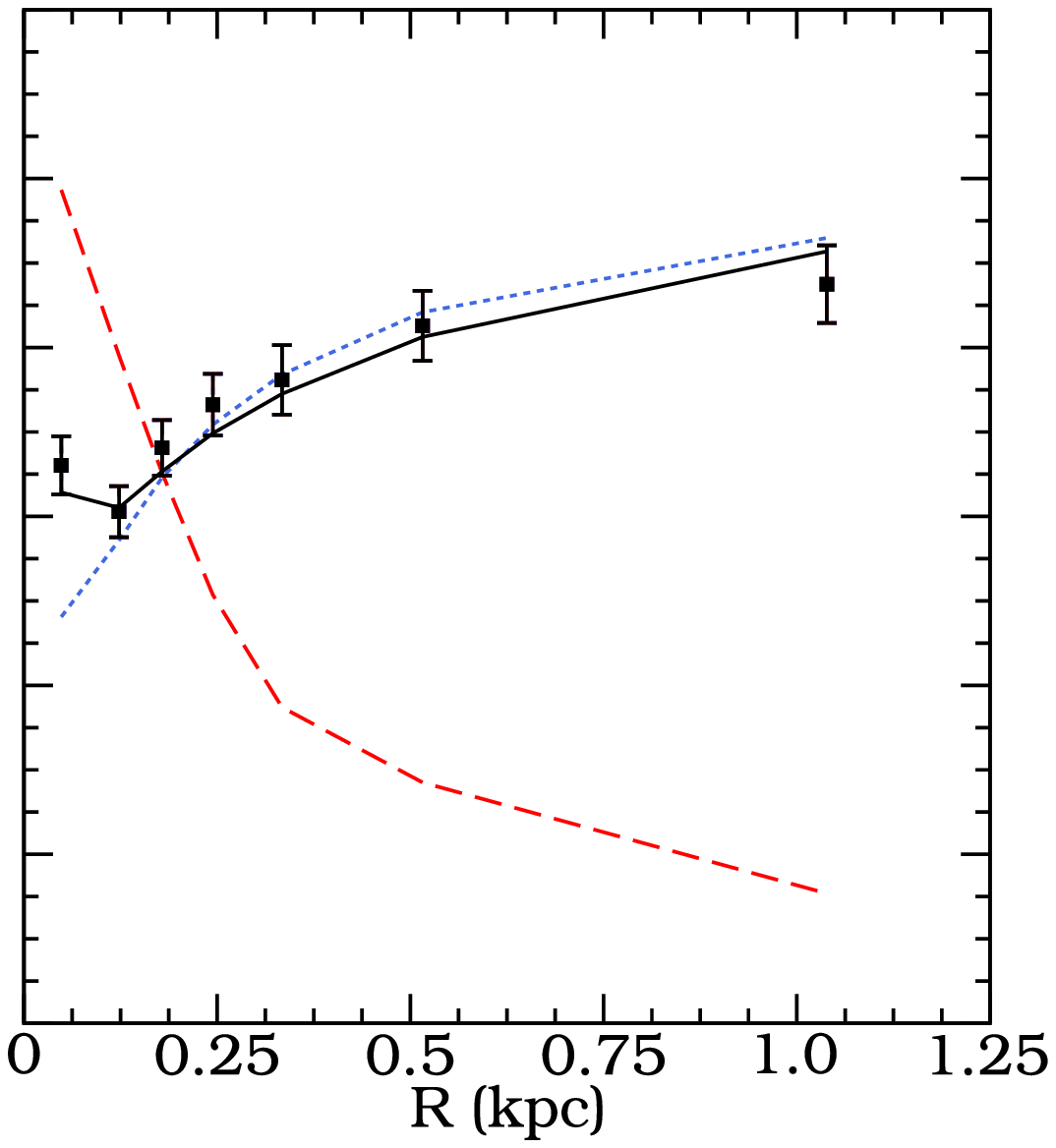} \\
\end{tabular}
\centerline{\includegraphics[scale=0.37]{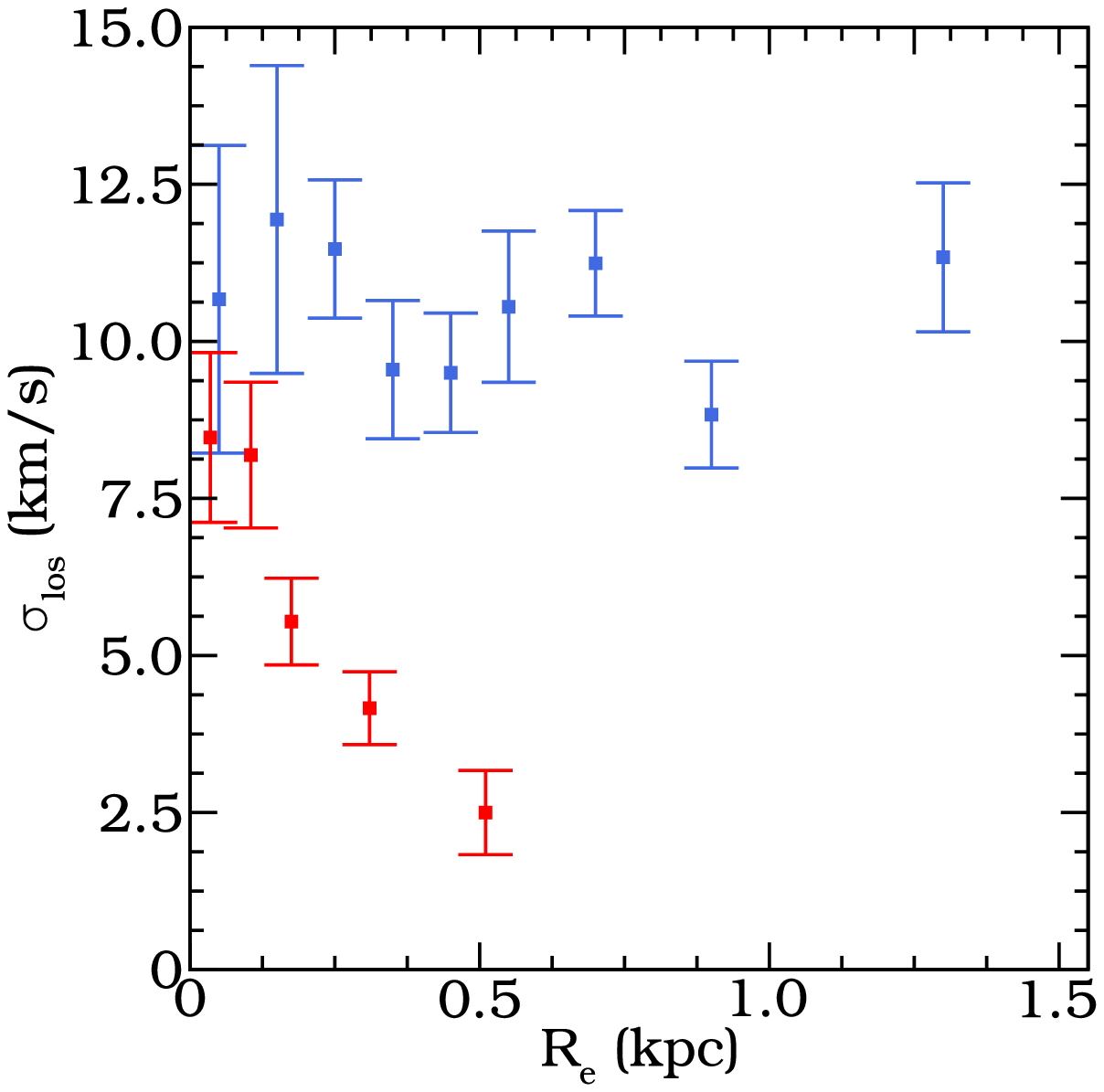}}
\vspace{3mm}
\caption{The black points with errorbars shown the velocity dispersion profile for Scl. The top
  four panels show the $\sigma_{\rm los}$  for each of our best fitting models,  separately for the low angular momentum
  (dashed red) and high angular momentum components (dotted blue) as identified in Fig. \ref{fig:scl-df}.
  The bottom panel shows the velocity dispersion profiles as function of elliptical radii for
  the metal rich and metal poor populations in Sculptor, using data from
  \citet{Battaglia2008ApJ} and also from
  \citet{Walker2009AJ....137.3100W} for the outer most bins (see the text for more details). \label{fig:vdisp} }
\end{figure}

\section{The nature of the bimodality}

From the distribution function, we can derive the line of sight
velocity as well as the surface brightness profiles for each
component.  This is useful in order to understand the link to the
cold/hotter kinematic or RHB/BHB populations known to exist in
Sculptor. In Fig.~\ref{fig:vdisp} we show the line of sight velocity
dispersion profiles for the two components separately, where each
panel corresponds to the best fit models shown in
Fig~\ref{fig:scl-df}.  The red component shows an increasing velocity
dispersion towards the centre and falls off rapidly with radius, while
the blue component has a flatter profile.

These trends are quite similar to the velocity dispersion profiles
shown at the bottom of the figure for the MR and MP components defined
as those stars with [Fe/H] $\ > -1.5$ dex
(red) and [Fe/H] $< -1.7$ dex (blue) in \citet{Battaglia2008ApJ}. In this figure we have also
added the data from \citet{Walker2009AJ....137.3100W} in the four
outermost bins for a fairer comparison to the data shown in the top panels and used for the Schwarzschild models. Note that the sharp [Fe/H] cuts used in this panel to define the two
populations do not guarantee the absence of cross-contamination,
especially in the inner regions, where the populations overlap
spatially. In fact, any misassignment in the center will tend to lower
the dispersion of the MR population, while increasing that of the
MP, as perhaps these comparisons suggest.

\newcommand{\scalelight}{0.37}
\begin{figure}
\begin{tabular}{c c}
\textbf{discrete, nfw} & \textbf{moments, nfw} \\
\includegraphics[scale=\scalelight]{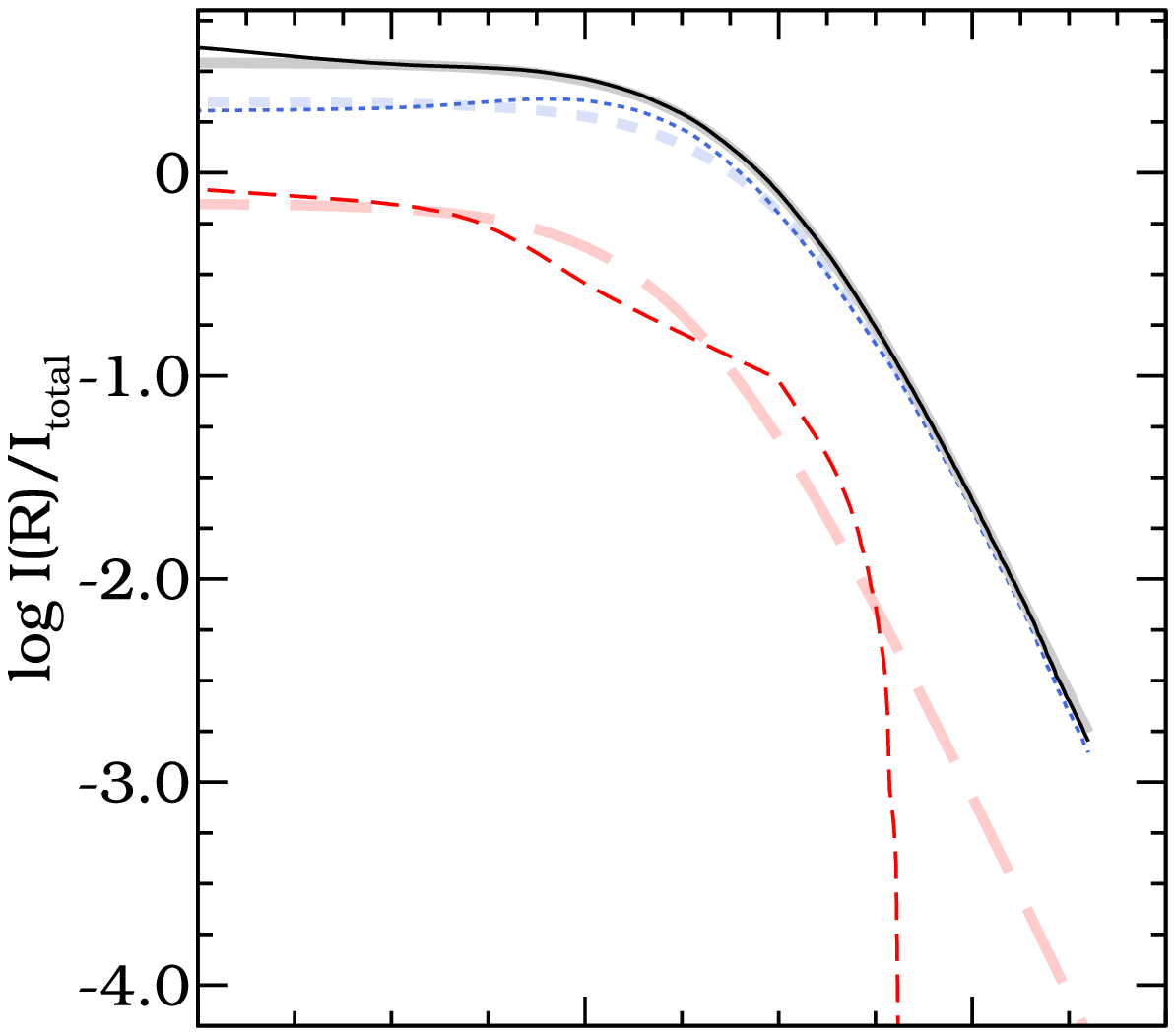}
& \includegraphics[scale=\scalelight]{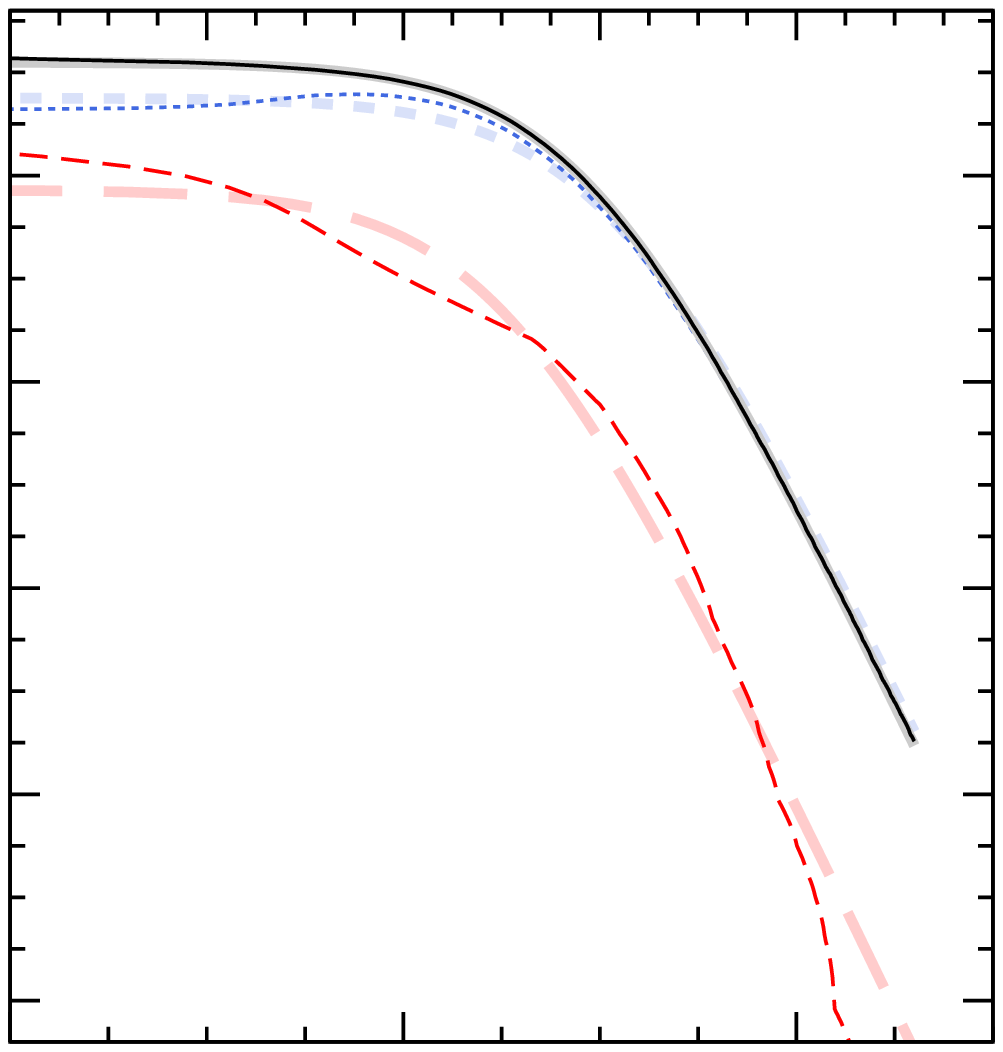}
\\
\textbf{moments, core13} & \textbf{moments, core24} \\
\includegraphics[scale=\scalelight]{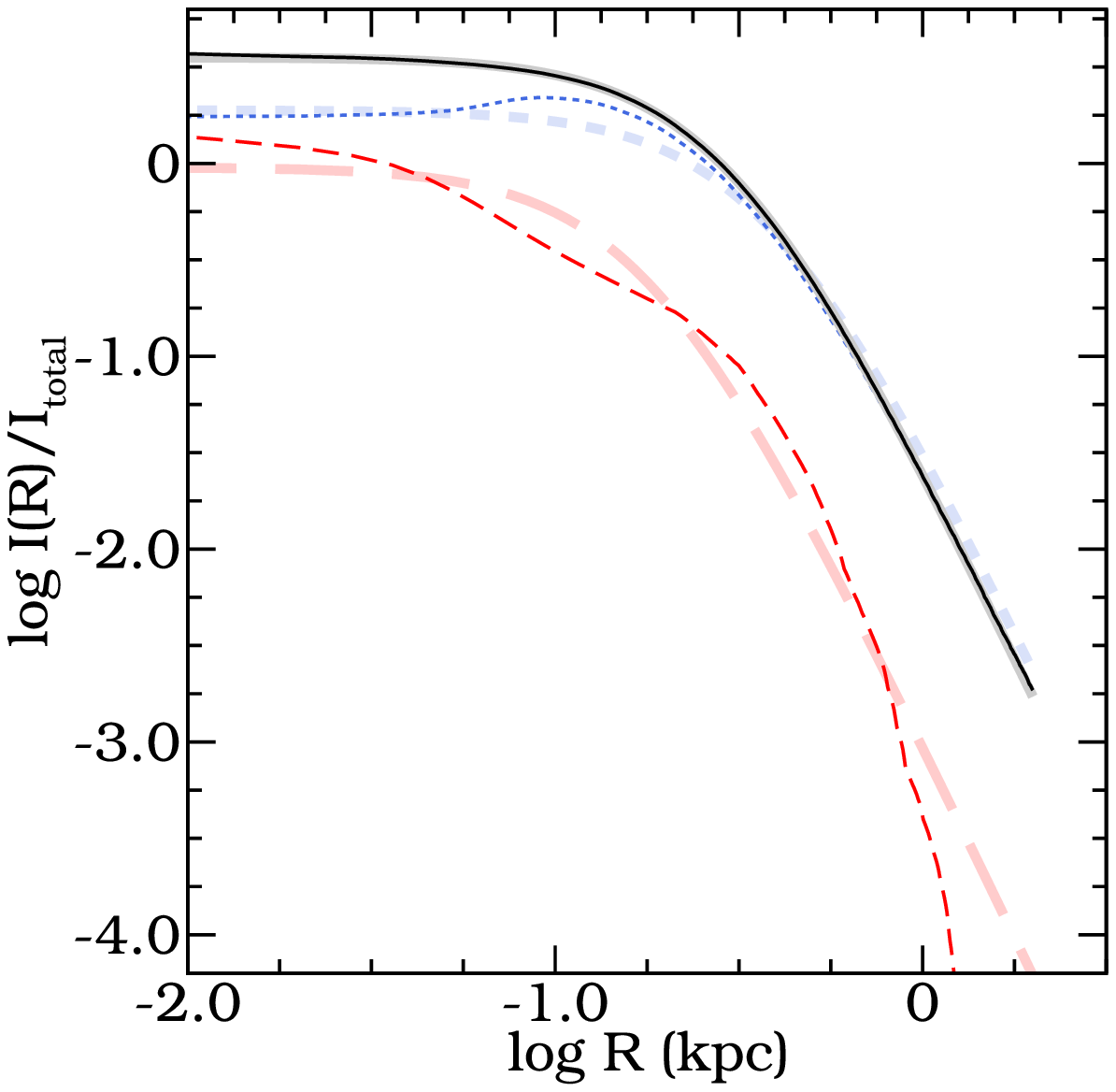}
& \includegraphics[scale=\scalelight]{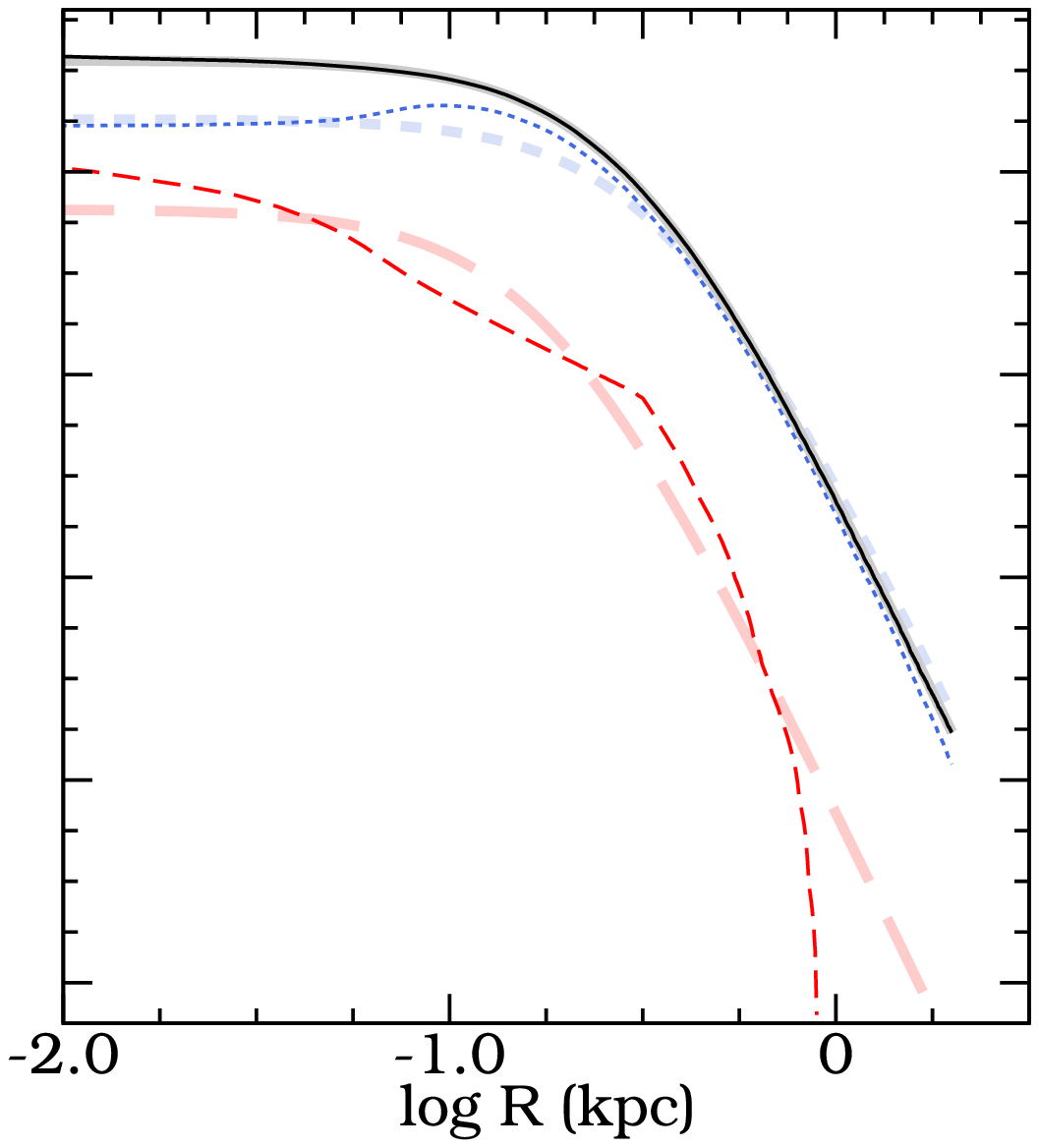}
\\
\end{tabular}
\caption{Density of the projected light distribution for the two
components, similar to Fig. \ref{fig:vdisp}. Solid black graph
indicates the total light distribution. Thicker and semi-transparant
lines indicate the best fitting Plummer profile.\label{fig:light}}
\end{figure}

In Fig.~\ref{fig:light} we show the projected light distribution for
each of the two identified components (in red and blue, for low and
high angular momenta respectively), as well as for the full system (in
black).  This figure clearly shows that the red component is more
centrally concentrated than the blue component for all models. This
can be quantified further by fitting a Plummer profile to each
component separately. The results are given in Table
\ref{tab:radii}. Note however, that the Plummer functional form does
not fit well the light distribution of the more concentrated, lower
angular momentum component (which is somewhat better fit with
an exponential).  Typically the ratio of Plummer scale radii is $b_{red}/b_{blue} \sim 0.5$,
which is comparable to the ratio determined by fitting Plummer
profiles to the RHB and BHB populations separately, which is $r=0.6$
according to \citet{BattagliaThesis}.  However, recall that the
photometric and even the spectroscopic decomposition do not
necessarily correspond to the dynamical components found by our
method. Nonetheless, the similarities are striking.

In terms of the velocity anisotropies of the two components detected,
we find the low angular momentum component to be radially anisotropic
in all cases considered. On the other hand, the high angular momentum
component is tangentially biased over most radii in all models, but it
shows larger model to model variations.  These reflect the differences
in the detailed form of the distribution function which can be seen in
Fig.~\ref{fig:scl-df}.

In the above analysis, parts of the distribution function were not
assigned to either of the two components (e.g. typically in the region
in between). None of the conclusions we have reached so far are very
sensitive to whether this intermediate mass is assigned to the red or
the blue components. We have found by respectively assigning this
  intermediate mass to the blue or red components, that the surface
  brightness profiles become smoother (i.e. less bumpy), but that the
  changes to the velocity dispersion profiles are very small ($< 1 km/s$). In
  particular the blue and red components remain clearly kinematically distinct in the central regions.
 The fractional masses for the intermediate component for each model can be
  found in the right columns of Table \ref{tab:radii}.

\begin{table}
\begin{tabular}{l|r r r | c c c}
 & $b_{red}$ & $b_{blue}$ &  $b_{total}$  & $f_{red}$ & $f_{int}$ & $f_{blue}$ \\
 \hline
nfw(discrete) & 0.19 & 0.34 & 0.3  & 0.080 & 0.109 & 0.812 \\
nfw           & 0.18 & 0.35 & 0.3  & 0.088 & 0.020 & 0.892 \\
core13        & 0.18 & 0.38 & 0.3 & 0.097 & 0.045 & 0.858 \\
core24        & 0.18 & 0.37 & 0.3 & 0.069 & 0.129 & 0.802
\end{tabular}
\caption{Plummer scale radii $b$ for the red (low angular momentum) and blue (high angular momentum) components, and for
the total light distribution. The last three
columns give the mass fractions associated to the red, blue and the intermediate components. }\label{tab:radii}
\end{table}

\section{Discussion and Conclusions}
\label{sec:conclusions}
We have shown that the distribution function of Sculptor for several best
fitting mass models is bimodal in energy and angular momemtum
space. The two components may be split in a low and high angular
momentum parts using the watershed method. The properties of the low and high
angular momentum components are similar to the metal rich and metal-poor
components respectively, known to be present in this galaxy, in terms
of their velocity dispersion profile and their light distribution \citep{Battaglia2008ApJ}.

This result is quite remarkable since we have not assumed at any point the existence of
multiple components in the Sculptor dwarf. It therefore suggests
that the metal-rich and metal-poor stars indeed are dynamically
distinct, and that Sculptor is not simply a system with a radial
gradient in stellar populations. This finding highlights the full power of the Schwarzschild's dynamical approach, and
would not have been possible if we had taken a parametric approach to model the
distribution function.

The fact that our models naturally recover the bimodality present in
Sculptor, for all dark matter profiles explored, and even for the NFW
form, would seem to be at odds with the results of \citet{Agnello} and \citet{Walker2011ApJ}.
In these works, Sculptor was modeled as a
two-component system with light distributions that followed similar
profiles. The use of the virial equations or the robust estimator of the mass at the half-light radii
of each component was used to argue that NFW profiles could be ruled out with high significance. Our
modeling however, shows this is not the case and why. The assumption of
similar light profiles appears to be crucial to reach those
conclusions, and would seem to introduce a systematic bias that none
of these works have taken into account. In the resulting
non-parametric dynamical models we have obtained, the light
distribution for the two components is quite different and cannot be parametrized well by Plummer profiles as Fig.~~\ref{fig:light} shows. Walker (2014, private communication) repeated the analysis of \citet{Walker2011ApJ}, but now allowing the metal rich profile to follow an exponential form. While still excluding a cusp with $p>98.4\%$ instead of $p>99.8\%$, it demonstrates the bias caused by the assumed parametric form of the light profiles.

It is natural to wonder whether other dwarfs also exhibit
multiple dynamical components.  Fornax would be a natural candidate
but its complex light distribution
\citep[e.g. the presence of shells, non-axisymmetries in the centre,][]{Battaglia2006AA} and the
hints of misaligned kinematics \citep[see e.g.][]{2012ApJ...756L...2A} require the use of a more general (non-spherical)
Schwarzschild modeling approach. On the other hand, it would be desirable to have
larger datasets for e.g. Carina and Sextans to be confident in the robustness of the analyses. More generally, an interesting
challenge will be to understand how such complex systems can form on the smallest galaxy scale, i.e. on the scale of the
dwarf spheroidals.

\section*{Acknowledgements}
AH and MB are grateful to NOVA for financial support. AH acknowledges financial support the European Research Council under ERC-Starting Grant
GALACTICA-240271.

\bibliographystyle{apj}

\end{document}